\def\be{\begin{equation}}
\def\ee{\end{equation}}
\def\bear{\begin{eqnarray}}
\def\eear{\end{eqnarray}}
\def\bdm{\begin{displaymath}}
\def\edm{\end{displaymath}}

\documentclass{article}

\usepackage{arxiv}
\usepackage{amsmath,amsfonts,amssymb}
\usepackage[utf8]{inputenc} 
\usepackage[T1]{fontenc}    
\usepackage{hyperref}       
\usepackage{url}            
\usepackage{booktabs}       
\usepackage{amsfonts}       
\usepackage{nicefrac}       
\usepackage{microtype}      
\usepackage{lipsum}
\usepackage{graphicx}
\usepackage[center]{caption}
\usepackage{enumerate}
\usepackage[inline]{enumitem}

\title{Skin lesion segmentation and classification using deep learning and handcrafted features }

\author{ Redha Ali \\
  Department of Electrical and Computer Engineering\\
  University of Dayton\\
  300 College Park, Dayton, Ohio 45469 \\
  \texttt{almahdir1@udayton.edu} \\
   \And
 Hussin K. Ragb \\
 Department of Electrical and Computer Engineering\\
 Christian Brothers University\\
  Memphis, Tennessee\\
  \texttt{hragb@cbu.edu} \\
  }

\begin{document}
\maketitle

\begin{abstract}
Accurate diagnostics of a skin lesion is a critical task in classification dermoscopic images.
In this research, we form a new type of image features, called hybrid features, which has stronger discrimination ability than single method features. This study involves a new technique where we inject the handcrafted features or feature transfer into the fully connected layer of Convolutional Neural Network (CNN) model during the training process. Based on our literature review until now, no study has examined or investigated the impact on classification performance by injecting the handcrafted features into the CNN model during the training process. In addition, we also investigated the impact of segmentation mask and its effect on the overall classification performance. Our model achieves an 92.3\% balanced multiclass accuracy, which is 6.8\% better than the typical single method classifier architecture for deep learning.
\end{abstract}


\section{Introduction}
Skin cancer is a cancer type that has a significant impact on society in the United States and across the world. The American Cancer Society’s report in the year 2019 estimated that 96,480 new cases of melanoma are expected (57,220 in male and 39,260 in female), and 7,230 people will die of melanoma (4,740 males and 2,490 female) \cite{american2019cancer}. Therefore, to decrease the number of deaths, scientists and researchers have made a tremendous effort to reduce the impact of skin cancer. An early stage of melanoma detection increases one’s survival rate (99\% if diagnosed in Stage I) in the United States. However, when the disease reaches the lymph nodes, the survival rate decreases to 63\% and if further decreases to 20\% when the disease spreads to distant organs~\cite{staff2018cancer,esteva2017dermatologist}.

To aid patients in spotting melanoma, dermatologists recommend the use of the ABCDE rule. 
The ABCD rule was introduced in \cite{friedman1985early} in 1985 as the ABCDE rule and then extended to the ABCDE rule in 2004 \cite{abbasi2004early}  and involves some key features of melanoma. 
However, the accuracy of unaided visual inspection is about 60\% \cite{kittler2002diagnostic}. 
The accurate diagnosis of the skin lesion is by taking a biopsy specimen where it enables the pathological analysis to differentiate between different types of the disease.  However, this type of analysis is not always possible, and it is both time and labor-intensive \cite{mahbod2019fusing}.
Recently, medical professionals have used dermoscopy, a new imaging technology that helps dermatologists to diagnose skin lesions more accurately than the unaided eye. Dermoscopy is a microscopy-based diagnostic method, and it is a non-invasive Imaging Techniques\cite{argenziano2002dermoscopy}.
Dermoscopy images generate a large amount of detailed images of skin cancer, and magnify the skin and eliminate surface reflection. However, the results of several studies have shown that an expert dermatologist can achieve a diagnostic accuracy between roughly 75\% to 84\% \cite{kittler2002diagnostic,vestergaard2008dermoscopy}. 
The risk of the wrong diagnosis can be increased with the lack of experience; this leads to interest in using artificial intelligence and machine learning to develop a Computer-Aided Diagnosis (CAD) system that provides a second opinion to help the dermatologist.

The architecture of a skin cancer CAD system typically consists of three stages: \begin{enumerate*}[label=(\roman*)] \item Lesion Segmentation, \item Attribute Detection, \item Disease Classification \end{enumerate*}. Our contribution focuses only on the first and third stages. The International Skin Imaging Collaboration (ISIC) is an academic and industry partnership designed to improve melanoma diagnosis, sponsored by the International Society for Digital Imaging of the Skin (ISDIS). The ISIC archive holds the most extensive publicly available dataset of dermoscopy images of skin lesions. The ISIC organization will benefit from the use of state-of-the-art machine learning technology.

Computer vision and machine learning have progressed rapidly and played an important role in the medical field, including medical image segmentation, detection, and classification \cite{razzak2018deep,ali2018leaf,ragb2016histogram,ragb2016, ragb2016multi}. Convolutional neural networks (CNN) \cite{lecun1998gradient} are a type of widely used deep artificial neural network. CNNs outperform the state-of-the-art in several computer vision applications \cite{simonyan2014very,brosch2014modeling}. CNN have also been presented in the field of medical image analysis \cite{simonyan2014very}. Several automated CAD system papers have been presented in the literature \cite{tran2021tmd, ali2019deep, li2018skin,ali2019skin,ragb2020deep,quang2017automatic, ali2020ensemble,hardie2018skin}. Moreover, various ensemble systems have proven very practical and versatile in medical image analysis domains and real-world applications \cite{ali2019fused, ali2019deep, ali2020ensemble, ragb2020deep, narayanan2019convolutional}. However, there is still an opportunity for further development in the accuracy of its diagnosis. 

In this research, we propose an algorithm for skin cancer segmentation and classification at a more treatable stage. Our proposed approach is computationally efficient and combines information from both deep learning and handcrafted features. We develop hybrid features, a new type of image feature, which has a more solid discrimination ability than single method features. This study involves a new technique that will inject the handcrafted features into the fully connected layer of the CNN model during the training process and enhance the segmentation model for better-handcrafted features.

\section{Materials and Methods}
\label{sec:headings}

\subsection{Dataset}

The International Skin Imaging Collaboration (ISIC) is an academic and industrial partnership designed to aid in melanoma detection, sponsored by the International Society for Digital Imaging of the Skin (ISDIS) \cite{li2018skin}. The ISIC archive owns the most comprehensive publicly available dataset of dermoscopy images of skin lesions \cite{tschandl2018ham10000}. In this study, we used the 2018 challenge dermoscopy databases to train and evaluate our proposed model. The segmentation dataset consists of 2,594 skin lesion images with corresponding ground truth for training. Also, there are around 100 samples for validation and 1,000 samples for testing. However, the ground truth masks for validation and testing images are not provided.  The training dataset for the classification task consists of 10015 images.  The images are 8-bit RGB dermoscopy images ranging from $540 \times 722$ to $4499 \times 6748$ pixels.

\subsection{Data pre-processing and augmentation}
\begin{figure}[!tbp]
  \centering
    \includegraphics[width=\textwidth]{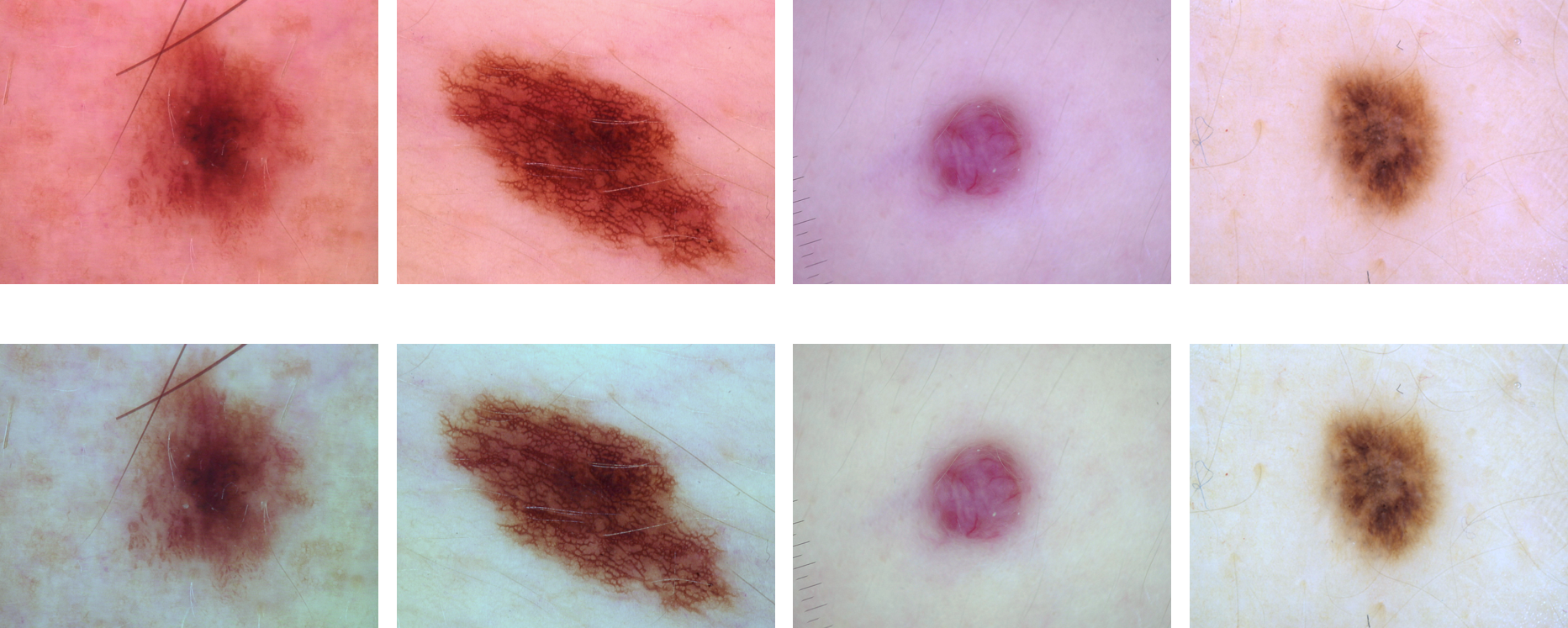}
    \caption{Examples of skin lesion images after preprocessing: (top) raw images,
(bottom) image obtained using color constancy.}
  \label{fig:fig1}
\end{figure}

The preprocessing step is performed using color constancy,
the Shades of Gray method, introduced by Finlayson \cite{finlayson2004shades}. We
applied this approach on all images during training
and testing as a preprocessing step with norm p=6. 
Preprocessing is crucial to normalize images across datasets
where acquiring the images uses an unknown light
source. Hence, we want to make all images appear identical
to colors under canonical light.
We used online data augmentations to prevent our model
from overfitting and to to improve performance and outcomes of our model. Also, data augmentations reduce the effect
of the small dataset. We performed data augmentation on
all datasets during training and testing. Data augmentation
included random crops, random rotation ($0-180^{\circ}$), vertical
and horizontal flips, and shear ($0-30^{\circ}$). All augmentation
hyperparameters were selected randomly. Figure \ref{fig:fig1} shows
examples of skin lesion images after preprocessing.
Finally, Each image is resized to fit the input of ResNet50\cite{zagoruyko2016wide}, and DenseNet201  \cite{huang2017densely} with an image input size of 224 x 224 pixels.

\subsection{Segmentation Model}
\label{sec:Segmentation}

\begin{figure}[!tbp]
  \centering
    \includegraphics[width=\textwidth]{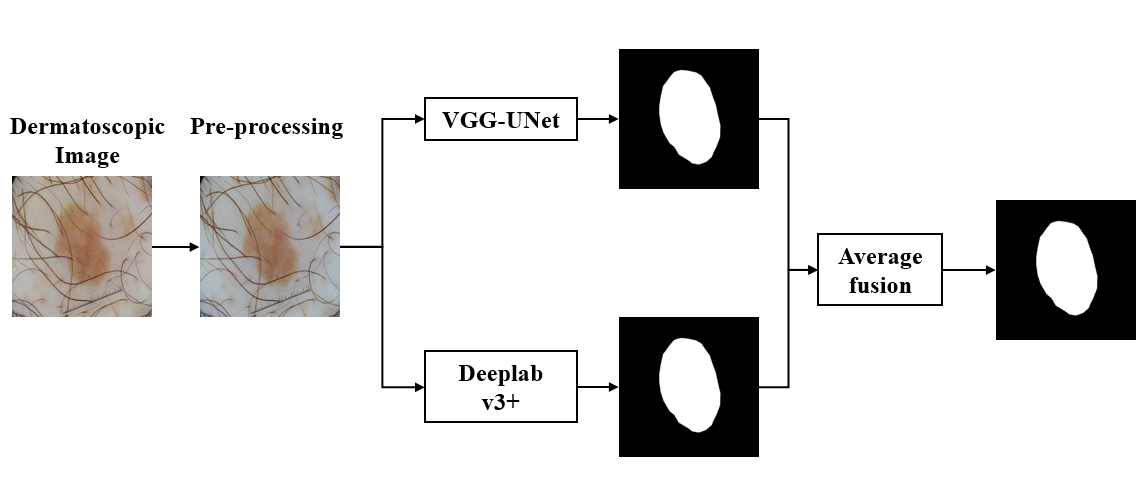}
    \caption{Illustration of our ensemble framework for lesion segmentation.}
  \label{fig:fig2}
\end{figure}

We utilizes an ensemble of two new deep learning architectures which introduced in \cite{ali2019deep}, including VGG19-UNet, and DeeplabV3+. The illustration of our ensemble structure for
lesion segmentation is shown in Fig. \ref{fig:fig2}.
The architecture details of VGG19-UNet and DeeplabV3+ described in \cite{ali2019deep}. 
The VGG19-UNet and DeeplabV3+ models were trained using Dice coefficient loss (also known as Srensen-Dice similarity coefficient). The Dice coefficient loss is based on the Srensen-Dice similarity coefficient for measuring overlap between two segmented images. In the ensemble technique, we employed an unweighted average where the probability matrices have been averaged to form the final predicated segmented mask for every dermoscopy image. The ensemble approach is shown in Fig. \ref{fig:fig2}. 

\subsection{Handcrafted Features}

For each detected and segmented lesion, the handcrafted features listed in \cite{messay2010new} are computed.
The features are computed from the RGB image along
with the lesion segmentation mask obtained using the method in segmentation model \ref{sec:Segmentation} above. The features
employed are similar to those used in \cite{messay2010new}. However, here they are computed for each of the three color channels and concatenated. The 200 geometric features are computed based on the shape and position information provided by the final segmentation mask.

\subsection{Classification Models}  
\begin{figure}[!tbp]
  \centering
    \includegraphics[width=\textwidth]{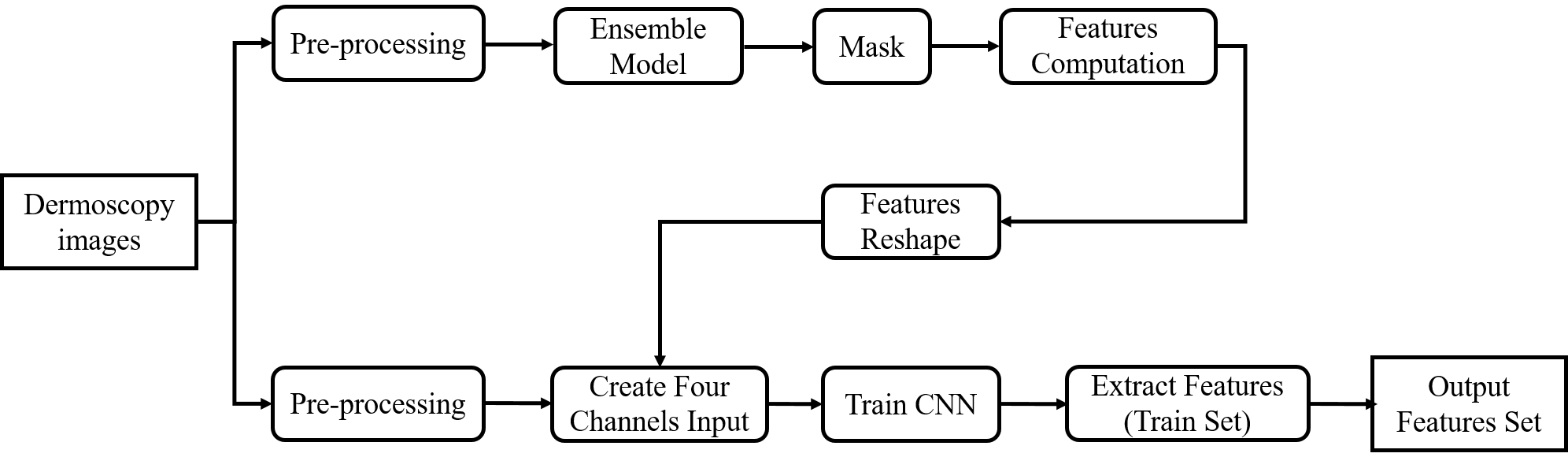}
    \caption{Illustration of our ensemble framework for lesion classification.}
  \label{fig:fig3}
\end{figure}

In this paper, we developed a novel and efficient deep learning training strategy to train a signal CNN model with both raw image and handcrafted features. The proposed hybrid model is illustrated in Fig. \ref {fig:fig3}.  The overall hybrid model consists of two core modules: (i) ensemble segmentation model and 200 handcrafted features. (ii) transfer learning using pre-trained CNN architecture with the minor modification allows the pre-trained CNN to utilize both raw image and handcrafted features.  To begin, we first segment all pre-processed dermoscopic images using our ensemble segmentation model to compute the 200 handcrafted features. We then use those computed features to fine-tune the pre-trained models along with the pre-processed dermoscopic images. Our model used two pre-trained models, including ResNet-50 and DenseNet-201. We concatenate the features that the last convolution layer has produced with handcrafted features in channel dimensions. Then, we inject the handcrafted features into the fully connected layers during the training and fine-tune the pre-trained models.  Finally, we extracted the features from both models and fed them to SVM for final predictions. 

\begin{table}[!tbp]
\centering
\caption{Network models}
\captionsetup{justification=centering}
\begin{tabular}{|c|c|c|c|}
\hline
Network     & Depth & Parameters (Millions) & Image Input Size  \\ \hline
ResNet50    & 50    & 25.6                  & 224 x 224         \\
DenseNet201 & 201   & 20                    & 224 x 224     \\ \hline   
\end{tabular}
\end{table}

\subsection{Evaluation Metrics}\label{metrics_section}

It is important to assess the efficacy of classification algorithms to aid in method comparisons, method selection, understanding system limitations, and to identify opportunities for future improvement.  The metrics we use as performance and efficiency metrics is balanced accuracy (${\bf BACC}$) specificity (${\bf SPEC}$), sensitivity (${\bf SENS}$), ${\bf Accuracy}$, and the area under the curve (AUC) for the receiver operating characteristic (ROC) curve. This metrics provide an objective quantitative picture of the efficacy of the systems tested.

The classification problem studied in this paper is multi-class classification.  Thus, each exemplar fits into one of four subsets:
\begin{enumerate}
  \item True positive (TP):  Interpreted as the classifier correctly predicts the positive sample.
  \item True negative (TN):  Interpreted as the classifier correctly predicts the negative sample.
  \item False positive (FP):  Interpreted as the classifier incorrectly predicts the positive sample.
  \item False negative (FN):  Interpreted as the classifier incorrectly predicts the negative sample.
\end{enumerate}
Based on the cardinality of these subsets, we can calculate the statistical quantities for each metric as follows:

\be
 {\bf BACC} =  \dfrac{{(TN/(TN+FP))+(TP/(TP+FN))}}{2},
\label{BACC_eq}
\ee

\be
 {\bf SPEC} =  \dfrac{{TN}}{(TN+FP)},
\label{SPEC_eq}
\ee

\be
 {\bf SENS} =  \dfrac{{TP}}{(TP+FN)}.
\label{SENS_eq}
\ee
and
\be
 {\bf Accuracy} =  \dfrac{{TP+TN}}{(TP+TN+TP+FN)}.
\label{Accuracy_eq}
\ee

On the other hand, the AUC metric\cite{bradley1997use} is the area under the ROC curve and it captures the degree of separability between classes. 
The higher AUC score represents a better model efficiency and vice versa.

\section{Results and Discussion}
As mentioned, the reported results are based on the 3004 test images of the ISIC 2018 skin lesion classification challenge.  Since the ground truth of the testing dataset was not provided by the challenge organizers, 
We randomly split the dataset into $70\%$ for training and $30\%$ for testing examples for each class. 
Then, we divided the training dataset into $90\%$ and $10\%$ for training and validation sets. 
Table \ref{table:table2} shows the hold-out validation distribution of the ISIC 2018 skin lesion dataset and the number of training and testing samples.

Table \ref{table:table3} shows the performance metrics for our approaches, including the proposed hybrid model and the ensemble of the hybrid models, with two baseline models for skin lesion classification using dermoscopic images. Note that the ensemble of the hybrid models obtained the highest BACC ($92.3\%$), outperforming the hybrid model and baseline models in this experiment. Furthermore, note that the hybrid model trained with handcrafted features outperforms the baseline models. Note that indicates the proposed hybrid training strategy is very effective and improved the balanced accuracy of ResNet-50 and DenseNet-201 by 3\% and 3.9\%, respectively.

\begin{table}[!tbp]
\centering
\caption{Shows the hold-out validation distribution of the dataset  and the number of training, validation, and testing examples}
\captionsetup{justification=centering}
\begin{tabular}{|c|c|c|c|} 
\hline
Disease Categories &  Image Input Size & No. of training set  & No. of testing set  \\ \hline
Actinic keratosis   &  224 x 224 & 229  & 98       \\
Basal cell carcinoma &  224 x 224 & 360 & 154    \\
Benign keratosis  &  224 x 224 & 769 & 330  \\
Dermatofibroma &  224 x 224 & 81 &  34  \\
Melanocytic nevus &  224 x 224 & 4694 & 2011  \\
Melanoma &  224 x 224 & 779 &  334  \\
Vascular lesion &  224 x 224 & 99 & 43  \\ \hline
Total & - &  7011 & 3004 \\ \hline
\end{tabular}
\label{table:table2}
\end{table}

\begin{table}[htp]
\caption{Comparison of the proposed model with the baseline models ResNet-50 and DenseNet-201 }
\centering
\begin{tabular}{|c|c|c|c|c|c|c|c|}
\hline
Models & BACC & SPEC & SENS & Accuracy & AUC  \\ \hline
ResNet-50 & 85.2\% & 96.1\% & 74.3\% & 96.3\% & 0.974  \\
DenseNet-201 & 85.5\% & 95.7\% & 75.2\% & 95.8\%  & 0.968 \\
Hybrid ResNet-50 & 88.2\% & 96.5\% & 79.9\% & 96.0\% & 0.974  \\
Hybrid DenseNet-201 & 89.4\% & 96.8\% & 82.0\% & 96.5\%  & 0.976 \\
Ensemble &  92.3\% & 97.1\% & 87.5\% & 95.4\%  & 0.977 \\ \hline
\end{tabular}
\label{table:table3}
\end{table}

In deep learning, the technique of combining multiple CNNs to form an ensemble is commonly used to improve performance. However,  the main contribution of this paper is to investigate the effect of injecting the handcrafted features to the pre-trained CNN during the training that achieves outstanding classification performance on the ISIC 2018 challenge dataset.  Moreover, our proposed ensemble hybrid training strategy showed that fusing two networks remarkably improves classification performance.

\section{Conclusion}
Early, affordable, and rapid detection of skin cancer is essential at the early stages to increase one’s survival rate. We successfully proposed an effective classifier for a dermoscopic image as a first-line triage tool to aid patients in spotting melanoma at early stages by training deep transfer learning models using our hybrid training method and combining their outputs into the ensemble. We show that our proposed method exceeds the baseline networks where transfer learning is used.  Further research can be conducted by testing new convolutional neural network architectures or with different ensemble techniques. 

\bibliographystyle{unsrt}  
\bibliography{references}  



\end{document}